\documentclass{vgtc}                          

\graphicspath{{figures/}{pictures/}{images/}{./}} 

\usepackage{times}                     

\usepackage{tabu}                      
\usepackage{booktabs}                  
\usepackage{lipsum}                    
\usepackage{mwe}                       
\usepackage{csquotes}

\usepackage{mathptmx}                  

\onlineid{0}

\vgtccategory{Research}

\vgtcinsertpkg

\title{Understanding User Needs for Injury Recovery with Augmented Reality}

\author{Jade Kandel\thanks{e-mail: kandelj@cs.unc.edu}\\ %
        \scriptsize University of North Carolina at Chapel Hill %
\and Sriya Kasumarthi\thanks{e-mail: kasri@unc.edu}\\ %
     \scriptsize University of North Carolina at Chapel Hill %
\and Danielle Albers Szafir \thanks{e-mail: danielle.szafir@cs.unc.edu }\\ %
    {\scriptsize University of North Carolina at Chapel Hill}}

\teaser{
    \centering
  \includegraphics[width=\linewidth]{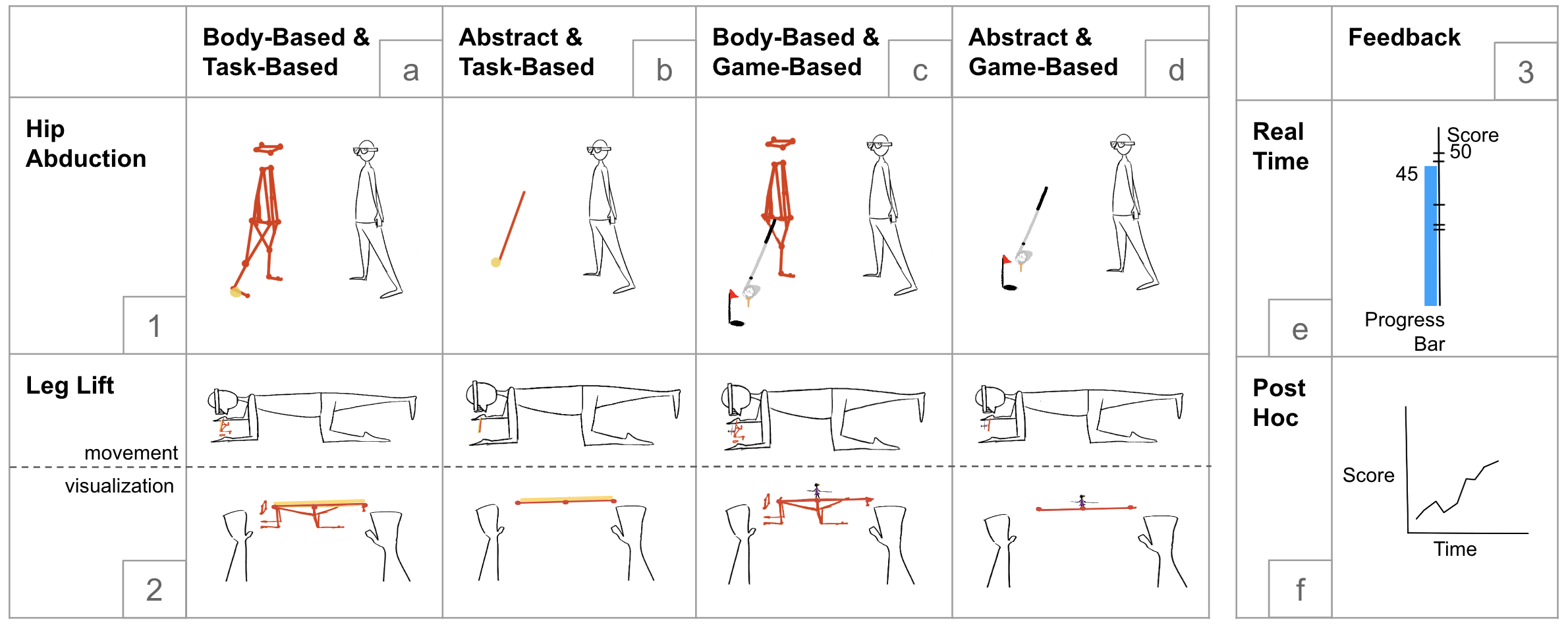}
  \caption{\textbf{Design Variable Examples} 
  Example visualizations for design variables. Visualizations are demonstrated with two exercise examples. Exercises consist of Hip Abudction (1) and Leg Lift (2) Visualizations consist of Body-based and Task Based (a), Abstract and Task Based (b), Body-based and Game-Based (c), and Abstract and Game-Based (d). Feedback methods (3) include Real-Time feedback (e) and Post-Hoc feedback (f)}
    \label{fig:examples}
}

\abstract{
    Physical therapy (PT) plays a crucial role in muscle injury recovery, but people struggle to adhere to 
    and perform PT exercises correctly from home. To support  challenges faced with in-home PT, augmented reality (AR) holds promise in enhancing patient's engagement and accuracy through immersive interactive visualizations. However, effectively leveraging AR requires a better understanding of patient needs during injury recovery. Through interviews with six individuals undergoing physical therapy, this paper introduces user-centered design considerations integrating AR and body motion data to enhance in-home PT for injury recovery. 
    Our findings identify key challenges and propose design variables for future body-based visualizations of body motion data for PT.
    
} 

\keywords{XR, visualization, health}

\begin{document}


\firstsection{Introduction}

\maketitle


Physical therapy (PT) plays an important role in recovery for people who have experienced an injury or are healing from surgery. PT aims to strengthen injured or weakened muscles, helping speed up the recovery process and lead to better health outcomes. While doing PT consistently and frequently at home significantly improves recovery, people may have a hard time knowing if they are doing the exercise correctly without a physical therapist present, which can lead to further injury and feeling insecure and nervous. People may also struggle with feeling motivated or engaged, resulting in neglecting to do the exercises as frequently as recommended. 

To address these challenges, leveraging body motion data—including real-time movement tracking and long-term progress metrics for muscle strength and flexibility—could significantly enhance patients' exercise accuracy, motivation, and overall engagement with their rehabilitation program. Augmented Reality (AR) offers significant advantages for presenting body motion data over traditional desktop videos or pictures. 
AR enhances the perception of spatial information such as depth, height, and size by integrating binocular cues and facilitating physical navigation \cite{marriott_immersive_2018,whitlock_graphical_nodate, kraus_value_2021, zacks_reading_1998}. This improved spatial perception is particularly crucial for analyzing the positioning of 3D body limbs accurately, where understanding depth and distance is essential. AR also enables body-based body motion data in the person's environment that is responsive and interactive, improving spatial understanding and engagement. We investigate how AR's strengths in presenting body motion data can be tailored to fit the needs for people recovering from sports injury.  

Previous research has explored using AR for rehabilitation \cite{doyle_base_2010,garcia_mobile_2014,ayoade_novel_2014}, but little research has investigated guiding exercise motion for sports
injury rehabilitation, especially for lower limbs \cite{butz_taxonomy_2022}. People who are injured and performing PT may face unique challenges such as limited mobility, limited body visibility, and slow recovery, which requires further insight and consideration when creating design guidelines. In this paper, we focus our attention on patient needs to propose user-centered and data-centered immersive design methods specific for injury recovery. 

We interviewed six individuals who in the past two years experienced an injury and were prescribed an at-home PT regime for recovery. We synthesized the feedback into key challenges patients experience when performing PT at home. Based on patient experiences and previous work, we propose immersive design variables and body motion data visualization techniques that may address these challenges. Specifically, we propose using tracked body motion data and visualizations to 1) show how to do the exercise correctly, 
2) increase engagement to improve adherence, and 3) present progress over time. Through analyzing patient challenges, and introducing key design techniques, we move towards future design guidelines that can significantly improve patient experiences and outcomes.


\section{Related Work}

Virtual and augmented reality are increasingly used in various domains for guiding body motion, such as improving dance performance \cite{laattala_wave_2024, anderson_youmove_2013}, cycling \cite{kaplan_towards_2018}, basketball free throws \cite{lin_towards_2021}, hand gestures \cite{freeman_shadowguides_2009}, yoga \cite{jo_flowar_2023}, and physical rehabilitation \cite{doyle_base_2010,garcia_mobile_2014,ayoade_novel_2014}. These systems situated visualizations with respect to the person or a digital copy of the person to provide intuitive, spatially-relevant feedback. Systems that compared AR feedback with video demonstration found that augmented feedback improved accuracy and engagement \cite{tang_physiohome_2015, anderson_youmove_2013, hoang_onebody_2016}. 

Visualization techniques for guiding motion use a range of representations, such as human model poses \cite{laattala_wave_2024, anderson_youmove_2013, jo_flowar_2023}, wedges \cite{tang_physiohome_2015}, or lines and circles \cite{doyle_base_2010-1}. While these systems focused on the execution of the exercise task, other systems incorporate game-based elements to increase engagement, such as catching gemstones with a butterfly in fantastical worlds \cite{elor_project_2019}, or disguising exercises as doing ball games, shotput, bowling, and balancing on a piece of wood in a river \cite{reilly_virtual_2021}. After surveying game-based exercise systems, Karaosmanoglu et al. proposed further exploration to understand which exercises and movements are optimized with XR exergames \cite{karaosmanoglu_born_2024}. Both game-based and task-based visualization elements show important potential for improving accuracy and adherence, which we incorporate in our design investigations to determine potential trade-offs and benefits.

While game-based and task-based methods focus on real-time body motion data, other body motion data related to muscle strength and flexibility may be useful for tracking progress across time for longer-term PT protocols. Systems such as ARKanoidAR \cite{cavalcanti_usability_2019}, interACTION \cite{bell_verification_2019}, and FruitNinja \cite{seyedebrahimi_brain_2019} add additional feedback using text, scores, graphs, or color. In this paper, we propose feedback techniques for showing progress, both real time and post-hoc, which may be especially relevant for slow injury recovery. 


\section{Challenges with Physical Therapy at Home}

\begin{figure}
  \includegraphics[width=\columnwidth]{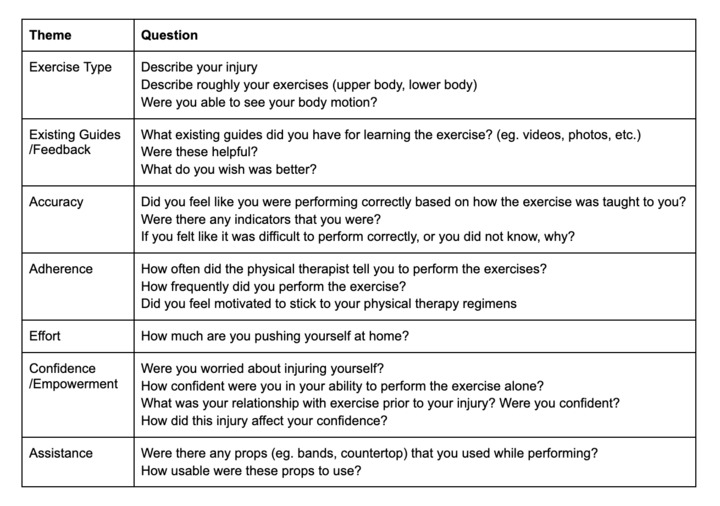}
  \caption{\textbf{Interview Questions:} The interview questions asked to our six participants, categorized in themes based on patient's experience.
  }
    \label{fig:questions}
\end{figure}

In order to create user-centric and data-centric designs, we need to better understand patient challenges and needs. We conducted 30 minute interviews with six participants (3 female and 3 male). All participants were recommended an at home PT regime after an injury and required at least a year of PT. Five participants experienced a lower-body injured from playing a sport, and the sixth had an upper-body injured from surgery. Of the five lower-body injuries, three participants experienced knee-related injuries, and 2 participants experienced hip-related injuries. 

We asked the participants about their experience doing PT to recover from their injury. We focused our questions around the added value that researchers aim to achieve using technology \cite{butz_taxonomy_2022}, such as accuracy, adherence, effort, and confidence (Figure \ref{fig:questions}). After conducting interviews, we coded responses and aggregated them into common themes reflecting opportunities where visualization may enhance PT procedures and outcomes, and highlighted patterns in responses. Despite the range of injuries and prescribed exercises, participants described similar challenges. Our findings reflect four core themes:

\textbf{Lack of visibility:} All participants said that their regime consisted of exercises that were either visible only in a mirror or not visible at all, especially for lower-body injury. For exercises requiring a mirror, three participants mentioned that not having a mirror or not having enough space in the room to see their body in the mirror made performing the exercise feel very difficult. One participant said ``I wish there was someone there that could actively correct me. Without a mirror in my room, it's really hard to tell. They say, \lq keep your hips level', but it's hard to tell if they are actually level." Participants preferred exercises that used a mirror or having a physical therapy present to watch them. 

\textbf{Lack of indicators for success:} Five participants said that they did not have any indication of success when alone at home, while the the sixth person had high self-reported proprioception, noting `` I have 10 years of yoga background and I know how to align the body and what the movements were designed to do." Two of the participants said that pictures were helpful for knowing how to do the exercise correctly, while the other four did not use photos or video as assistance, and  ``preferred to rely on memory than to have to pay attention to a picture."  One participant noted that the only possible indication of success was that ``my leg was getting stronger over time." Another said that ``The two (types of) feedback were visuals (from mirror) and pain. If there was pain, it was good... I know if it's a good pain because I go to a  physical therapist to test my limits and get feedback." This participant relied heavily on memory to remember how the exercise should look in the mirror and what the pain should feel like. Another participant said they knew if they were doing the exercise correctly ``based on vibes, nothing exact," and remarked how a lack of indicators for success affected their adherence because ``not being able to know if I was doing the exercise correctly made me not want to exercise (at home), so I waited to do physical therapy when the physical therapist was around to correct me." In summary, while having visibility in the mirror made it easier to know what their body was doing, there was still a lack of exact visual and physical indication that participants were aligning their body correctly, forcing them to rely on their memory and intuition. 

\textbf{Program adherence:} All participants were first asked by physical therapists to perform exercises every day for a couple months, and then the frequency was decreased over time as the body got stronger and improved. Two of the participants adhered to the program and did the exercises as frequently as recommended, but the other four participants did not. One participant explained their lack of adherence, ``I have other priorities. I am tired. I am busy. It takes so long." Two participants described physical therapy exercise as ``boring." Another participant with a very busy work schedule said ``I was not the type of person who had the time or the full motivation to get back to normal because that is not what my life is based on.” So while all participants acknowledged the necessity of physical therapy for recovery, four of them found it challenging to prioritize due to external factors or boredom. 

\textbf{Unable to see progress:} All participants mentioned that recovery was slow, and it took a long time to see any progress. One participant said, 
\begin{displayquote}
``Its not a perfect linear progression of improvement, there are bad days... I would like to know what I'm going to achieve by what time based on what I do, but it's hard because everyone reacts differently. I was feeling impatient. You don't know if you are going to heal, it's a psychological battle. The inability to do what you want to do is very challenging." 
\end{displayquote}

Another participant said ``I wasn't seeing progress. Speed of progress varies and is very personal." Recognizing the importance of focusing on the present, one participant said ``You need to be patient with your body. I cannot stress enough the importance of listening to your body, there should be strain but no pain. Full extension is the end goal."  Due to long injury recovery and the inability to see progress, people had to put effort into staying patient, trusting the process, and not losing motivation. 

\section{Design Variables}
Based on our interview findings, we identify key design variables for incorporating AR and motion data in injury rehabilitation. These variables, rooted in challenges articulated by PT patients, provide critical decision points for future visualization systems supporting PT. Figure \ref{fig:examples} illustrates these design variables using two limited-visibility, lower body PT exercises: hip abduction and leg lift.
The hip abduction exercise (Figure \ref{fig:examples}.a) involves the patient lifting one leg laterally away from their body. In the leg lift exercise (Figure \ref{fig:examples}.b), the patient starts on all fours, then elevates and holds one leg to form a straight line with their back. Our design variables address three primary aspects: body-based vs. abstract encodings, game-based vs. task-based designs, and real-time vs. post-hoc feedback. 

\textbf{Body-based vs. Abstract Encodings:}
Given that patients lack visibility of the motion and indication of success, using 
encodings that demonstrate accurate motion 
may help patients remember and execute their exercises correctly. Body-based encodings
visualize motion data
on the person's body or 
corresponding digital twin. Depending on the exercise, the digital twin may be life size and face the participant (Figure \ref{fig:examples}.1a and Figure \ref{fig:examples}.1c), or miniature (Figure \ref{fig:examples}.1b and Figure \ref{fig:examples}.1d). Abstract encodings simplify the visualization by using shapes without a body to cue motion (Figure \ref{fig:examples}.b and Figure \ref{fig:examples}.d). While the body-based encodings provides additional spatial context 
with the body present, abstract encodings are simpler which may improve usability and decrease cognitive load. 

\textbf{Game-based vs. Task-based Designs:} Given that patients feel fatigued and unmotivated, creating fun AR experiences could improve adherence by using a narrative or pictorial designs to disguise exercises as movements for completing a game-based goal. Game-based goals may include hitting a golf ball with a golf club into a hole (Figure \ref{fig:examples}.1c and Figure \ref{fig:examples}.1d), or keeping an acrobat balanced on a tight rope (Figure \ref{fig:examples}.2c and Figure \ref{fig:examples}.2d). Task-based designs do not disguise the motion and are direct in leading the person to complete the exercise task. Task-based visualizations may involve touching a target yellow ball (Figure \ref{fig:examples}.1a and Figure \ref{fig:examples}.1b), or aligning with a target yellow line (Figure \ref{fig:examples}.2a and Figure \ref{fig:examples}.2b). While game-based designs may be more motivating and fun, they may also be considered distracting, tiring, and potentially harder to cue proper motion compared to more direct task-based approaches. 

\textbf{Real-time vs Post-Hoc Feedback:} Tracking progress across time can help people struggling with how long recovery takes and feel more motivated by seeing their progress. Exercise progress tracking can utilize body motion data tailored to specific goals, such as measuring the proximity of body parts to target positions, counting successful repetitions, or timing the duration of held poses. Feedback to communicate this data can consist of high scores, graphs, and other statistical metrics demonstrating progress across time. Real-time feedback shows high scores and progress in real time (Figure \ref{fig:examples}.3e), while post-hoc shows progress across time after the exercise (Figure \ref{fig:examples}.3f). While real-time feedback may motivate people to try to push themselves to beat their previous high scores or feel a sense of accomplishment in the moment, people may push themselves too hard which may lead to injury. While post-hoc addresses this concern by providing a space after the exercise for analysis and processing, it may not be as satisfying or motivating to see the information after the exercise is complete.

\section{Exploratory Prototype}

\begin{figure}
  \includegraphics[width=\linewidth]{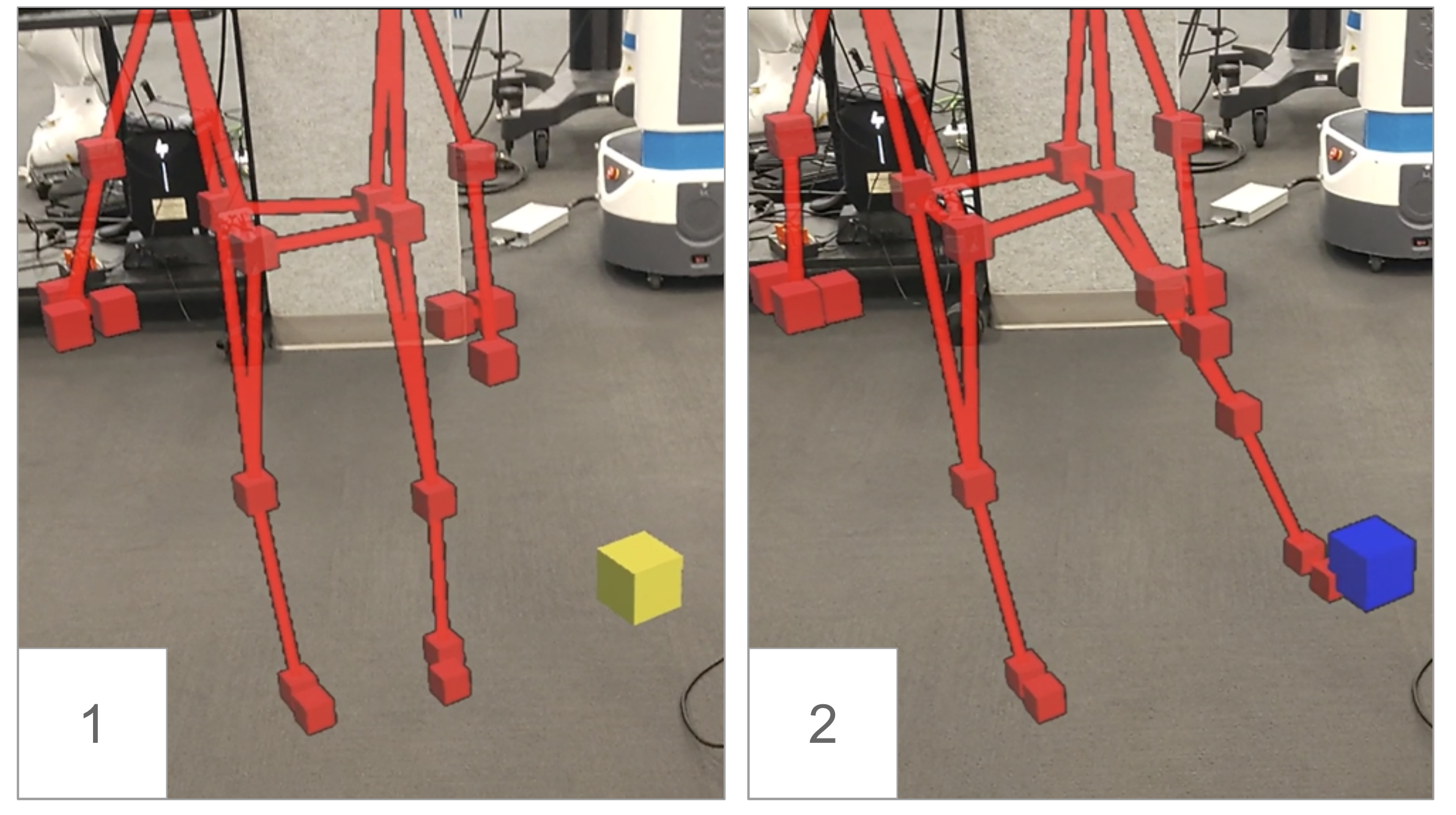}
  \caption{\textbf{Exploratory Prototype:} A visualization with the red skeleton digital twin and task-based target for hip abduction exercise before toughing the target (1) and after touching the target (2). Pictures taken in a HoloLens 2 headset.}
    \label{fig:pilot}
\end{figure}

Using a HoloLens 2 headset, Unity, and Vicon Nexus for motion capture, we implemented a body-based and task-based prototyped design in a preliminary system (Figure \ref{fig:examples}.1a). We were particularly interested in comparing two techniques for body-based visualizations: those situated with respect to the person's actual body and those associated with a digital twin. We conducted a pilot study with three participants, presenting them with two distinct visualizations: a target ball situated next to their own body and a target ball situated next to a red skeleton digital twin (Figure \ref{fig:pilot}). The red skeleton digital twin faced the participant and mirrored their real-time movements. In both scenarios, when the participant successfully touched the target ball with their left ankle, the ball turned blue to provide visual feedback. Key findings from the pilot study revealed that participants found it easier to look straight ahead and interact with the red skeleton digital twin, compared to looking down at their own body to interact with the visualizations directly. The red skeleton was not perceived as confusing or distracting. Instead, participants found it fun and engaging. Participants were not bored by the simple target and enjoyed turning it from yellow to blue. These preliminary results indicate that the digital twin approach may offer advantages in terms of user comfort and engagement, and that task-based visualizations could be engaging.  

\section{Discussion}

Our designs and prototype showcase visualization techniques aimed at enhancing in-home PT experiences. While this paper primarily addresses the patient's interaction with these visualizations, we envision future systems that seamlessly integrate physical therapists into the process. Currently, physical therapists' insights into patient recovery are constrained by brief clinical appointments and patients' often imperfect recollections. By leveraging body motion data collection and communication, we can enable physical therapists to remotely monitor patient progress and dynamically adjust targets and goals based on viewed data \cite{kandel_pd-insighter_2024}. Recognizing that each individual's healing process is unique, the integration of clinicians' expertise and analysis can facilitate the creation of personalized exercise regimens. This approach allows for the tailoring of rehabilitation goals to each patient's specific needs and progress, potentially improving outcomes and patient engagement. Future research could explore the most effective ways to present this data to therapists and how to incorporate their insights into the patient's AR-assisted exercises.

While this paper focuses on design-related aspects of in-home PT, the success of such systems depends on several factors. For example, significant advancements in motion capture technology and hardware accessibility are necessary. Systems like the Vicon Nexus offer highly accurate body motion capture, but the system's size, complexity, and cost makes it impractical for home use. More affordable and compact solutions such as the Kinect system, while more suitable for home environments, often lack the precision required for effective therapy monitoring. Recent progress in egocentric motion capture technology indicates that accurate body motion tracking could be achieved in the future using wearable devices and downward-facing cameras \cite{zhang_reconstruction_2023}, potentially providing a lightweight, non-intrusive solution for home-based motion capture. 
In addition, current AR headsets can be cumbersome and uncomfortable for extended use, particularly during physical exercises. The creation of lighter, more ergonomic headsets is essential to ensure patient comfort and compliance. By addressing these challenges, we can move closer to AR and motion capture systems that collect and present motion data effectively and integrate seamlessly into patients' daily lives.

Given the needs of people recovering from injury, many potential trade-offs must be considered in designing successful visual analytics approaches. For example, while we aim to make the exercises engaging and fun to increase adherence, it is also crucial that participants perform the physical therapy exercises correctly. Patients need to push themselves to strengthen their muscles and reach target goals, but not to the point of causing injury. Future studies can help us better understand these trade-offs and explore potential situated visualization techniques that balance multiple goals effectively.

\section{Conclusion}
AR has the potential to significantly improve patient's experience with at-home physical therapy. However, doing so requires understanding of the challenges patient face, and further investigation into the trade-offs of different visualization techniques. In this paper, we interviewed six people who did at-home physical therapy to recover from an injury. From their described challenges, we present design variables and the potential trade-offs and benefits for using these methods to cue motion. Through immersive visualizations and body motion data, we work towards improving at-home injury recovery by making the experience more engaging and safe.

\acknowledgments{
The authors wish to thank Henry Fuchs, Daniel Szafir, Michael Lewek, Spiros Tsalikis, and Jim Mahaney. This work was supported by
the National Institutes of Health Award 1R01HD111074-01 and NSF IIS-2320920.}

\bibliographystyle{abbrv-doi}

\bibliography{template}
\end{document}